\documentclass[pdflatex,sn-mathphys-num,iicol]{sn-jnl}

\usepackage{graphicx}%
\usepackage{multirow}%
\usepackage{amsmath,amssymb,amsfonts}%
\usepackage{amsthm}%
\usepackage{mathrsfs}%
\usepackage[title]{appendix}%
\usepackage{xcolor}%
\usepackage{textcomp}%
\usepackage{manyfoot}%
\usepackage{booktabs}%
\usepackage{algorithm}%
\usepackage{algorithmicx}%
\usepackage{algpseudocode}%
\usepackage{listings}%

\theoremstyle{thmstyleone}%
\theoremstyle{thmstyletwo}%

\theoremstyle{thmstylethree}%

\begin{document}

\title[Article Title]{Relativistic Modeling for Solid Earth Tide Estimation via Space-to-Ground Clock Comparison}

\author[1]{Qin Li}

\author[1]{Wei-Hang Sun}

\author*[1]{Yu-Jie Tan}\email{yjtan@hust.edu.cn}

\author[2]{Cheng-Gang Qin}

\author[1]{Jun Ke}

\author[3,4]{Xiang-Pei Liu}

\author*[3,4]{Han-Ning Dai}\email{daihan@ustc.edu.cn}

\author*[5]{Tong Liu}\email{liutong2021@csu.ac.cn}

\author*[1]{Cheng-Gang Shao}\email{cgshao@hust.edu.cn}

\affil[1]{National Gravitation Laboratory, MOE Key Laboratory of Fundamental Physical Quantities Measurement, and School of Physics, Huazhong University of Science and Technology, Wuhan 430074, People's Republic of China}

\affil[2]{MOE Key Laboratory of TianQin Mission, TianQin Research Center for Gravitational Physics \& School of Physics and Astronomy, Frontiers Science Center for TianQin, CNSA Research Center for Gravitational Waves, Sun Yat-sen University (Zhuhai Campus), Zhuhai 519082, China}

\affil[3]{Hefei National Research Center for Physical Sciences at the Microscale and School of Physical Sciences, University of Science and Technology of China, Hefei 230026, China}

\affil[4]{Hefei National Laboratory, University of Science and Technology of China, Hefei 230088, China}

\affil[5]{Technology and Engineering Center for Space Utilization, Chinese Academy of Sciences, Beijing 100094, China}

\abstract{With the rapid development of modern atomic clock technology, their unprecedented precision elevates them from timekeeping tools to gravitational potential sensors, thereby fostering the highly interdisciplinary field of Relativistic Geodesy. Given the potential for high-precision clock networks to detect periodic gravitational variations, it is imperative to assess their capability to invert solid Earth tide parameters via space-to-ground links in the presence of complex observational noise. To this end, we incorporate Earth's gravitational potential, direct lunisolar tidal potentials, and solid Earth tide effects into a high-precision relativistic framework for space-to-ground clock comparisons. By employing a three-link Doppler cancellation configuration to isolate the target signal, we perform numerical simulations for an inclined geosynchronous orbit satellite to analyze the effects of clock instability and colored precise orbit determination errors on parameter extraction. Our findings reveal that while high orbital altitudes cause severe collinearity between individual Love numbers, an effective parameter combining the $h_2$ and $k_2$ Love numbers successfully converges to a stable estimate within a 30-day continuous observation window. Furthermore, sensitivity analysis demonstrates that extraction accuracy is currently limited by clock stability rather than radial precise orbit determination errors.}


\maketitle

\section{Introduction}
\label{Introduction}
With the rapid development of modern atomic clock technology, the stability and uncertainty of time and frequency measurements have reached unprecedented levels. Currently, optical clocks based on optical lattices and single ions have demonstrated frequency uncertainties better than $1 \times {10^{ - 18}}$ in laboratory environments \cite{jiaImprovedSystematicEvaluation2026,liuZeroDeadTimeStrontiumLattice2025,marshallHighStabilitySingleIonClock2025a,yangClockPrecisionStandard2025}. This unprecedented precision elevates atomic clocks from merely precise timekeeping tools to a novel class of gravitational potential sensors, thereby catalyzing the emergence of Relativistic Geodesy as a highly interdisciplinary field \cite{mullerHighPerformanceClocks2017,mcgrewAtomicClockPerformance2018}.

According to the equivalence principle of general relativity, a gravitational field is locally equivalent to an accelerating reference frame. In an accelerating reference frame, clocks at different locations run at different rates due to their differing states of motion. It follows that clocks located at different points in a gravitational potential run at different rates, giving rise to the gravitational redshift effect. Specifically, a clock situated closer to a gravitational mass ticks slower than one located further away. In the weak-field approximation, the relationship between the fractional frequency shift and the potential difference can be expressed as $\Delta f/{f_0} \approx \Delta W/{c^2}$. For a frequency measurement precision of ${10^{ - 18}}$, the corresponding gravitational potential sensitivity is equivalent to a geopotential height difference of approximately 1 centimeter on the Earth's surface \cite{mehlstaublerAtomicClocksGeodesy2018,mcgrewAtomicClockPerformance2018}. This implies that high-precision clock comparisons can directly sense the fine structure of the Earth's gravitational field and its temporal variations.

Solid Earth tides refer to the periodic elastic deformations of the Earth mainly caused by the lunisolar tidal forces. These deformations induce vertical and horizontal displacements at ground stations and redistribute the Earth's internal mass, which in turn causes temporal variations in the local gravitational potential \cite{melchiorTidesPlanetEarth1983}. Traditionally, Earth tide parameters, such as Love numbers, have been determined using superconducting gravimeters, tiltmeters, or space geodesy techniques (e.g., VLBI, SLR, GNSS) \cite{schubertTreatiseGeophysicsGeodesy2007,neumeyerSuperconductingGravimetry2010}. While these established methods have achieved remarkable success, they primarily rely on measuring either the spatial derivatives of the potential (e.g., local gravitational acceleration via gravimeters) or the geometric manifestations of tidal deformations (e.g., surface displacements via GNSS and SLR). In contrast, high-precision atomic clocks offer a fundamentally different observational paradigm by directly sensing the gravitational potential itself. Consequently, the proposed clock comparison method serves as a completely independent and highly complementary observable. By providing direct measurements of the tidal potential variations, clock comparisons can help cross-check existing physical models, potentially decouple systematic errors inherent to purely geometric or gravimetric observations, and contribute to future joint inversions for a more comprehensive understanding of Earth's internal dynamics. Theoretical studies and preliminary experiments indicate that high-performance optical clock networks possess the capability to directly detect these periodic gravitational potential variations induced by Earth tides \cite{qinTidalClockEffects2020,qinTidalEffectsClock2023,qinRelativisticTidalEffects2019,lionDeterminationHighSpatial2017,mcgrewAtomicClockPerformance2018,bondarescuGroundbasedOpticalAtomic2015,zhangLunarTidalResponseinduced2026,zhangInfluenceLunarTidal2026}.

Compared to localized optical fiber-linked clock comparisons, space-to-ground or inter-satellite clock comparisons via satellite links offer significant advantages, including global coverage and the ability to span oceans and complex terrains. With the advancement of space-borne atomic clock projects (e.g., ACES/PHARAO) and the performance improvements of clocks onboard high-altitude satellites (such as GNSS), retrieving geophysical parameters via high-precision space-to-ground frequency links has become feasible \cite{delvaAtomicClocksNew2013,cacciapuotiSpaceClocksFundamental2009a}. Nevertheless, in long-baseline space-to-ground comparisons, the extracted signal is inextricably coupled with satellite orbit errors, atmospheric propagation delays, and other relativistic effects. A critical outstanding challenge is effectively isolating the target tidal gravitational potential from complex observational noise to accurately invert the Earth's elastic response parameters. Addressing this challenge requires advanced error-suppression techniques combined with a high-fidelity theoretical framework.

Within the framework of General Relativity, this paper incorporates the Earth's gravitational potential, tidal potential, and solid Earth tide effects into a high-precision physical model for space-to-ground clock comparisons. Taking the clock comparison between Inclined Geosynchronous Orbit (IGSO) satellites and ground stations as an example, we numerically simulate the impact of satellite orbit errors and clock noise on signal extraction via a three-link Doppler cancellation configuration. This approach enables us to investigate the feasibility and accuracy of using long-baseline clock comparison data to invert solid Earth tide parameters, aiming to provide a theoretical foundation and methodological reference for monitoring Earth's geodynamics using future space-based time and frequency networks.

The paper is organized as follows. Section \ref{theory} establishes the theoretical framework, detailing the relativistic clock comparison model, the three-link Doppler compensation scheme, and the solid Earth tide model. Section \ref{noisemodel} outlines the noise models, specifically characterizing clock instability and satellite precise orbit determination (POD) errors. Section \ref{simulation} presents the numerical simulation experiment, detailing the signal extraction methodology, parameter estimation results, and a sensitivity analysis. Finally, we give our conclusions in Section \ref{conclusion}.

\section{Theoretical Framework}
\label{theory}
\subsection{Relativistic Clock Comparison Model}
\label{clockcomparisonmodel}
High-precision frequency comparisons between satellites and ground stations must be treated within a four-dimensional relativistic spacetime framework. To ensure mathematical and physical consistency across all dynamical equations, the entire theoretical calculation and numerical analysis are strictly carried out in the Geocentric Celestial Reference System (GCRS), which serves as the fundamental quasi-inertial frame. The International Terrestrial Reference System (ITRS) is utilized to define the initial geodetic coordinates of the ground stations and to evaluate the Earth’s gravitational field models. Specifically, the kinematic state vectors of the ground stations (positions and velocities) are explicitly transformed from the ITRS to the GCRS using standard Earth rotation and transformation matrices. Meanwhile, the scalar gravitational potential $W$, being an invariant scalar under spatial coordinate rotations, is evaluated directly using the station coordinates and the gravity-field model within the ITRS, and subsequently incorporated into the GCRS spacetime metric. Under the weak-field approximation (neglecting higher-order relativistic terms related to Earth's rotation), the spacetime metric near the Earth can be written up to $O({c^{ - 3}})$ precision as \cite{soffelIAU2000Resolutions2003}
\begin{equation}
	\begin{aligned}
		g_{00} &= -1 + \frac{2W}{c^2} + O(c^{-4}) \\
		g_{0i} &= 0 \\
		g_{ij} &= \left(1 + \frac{2W}{c^2}\right)\delta_{ij} + O(c^{-4}).
	\end{aligned}
\end{equation}
The above metric neglects the frame-dragging effect (Lense-Thirring effect) induced by the Earth's angular momentum due to its negligible magnitude. The potential term $W$ includes the Earth's gravitational potential $W_E$ and external celestial potentials $W_{ext}$. In our calculations, external potentials are approximated as tidal potentials $W_{tide}$, omitting the significantly smaller inertial terms.

According to General Relativity, the proper time $\tau$ recorded by a moving ideal clock relates to the coordinate time $t$ (here taken as Geocentric Coordinate Time, TCG) along its trajectory. For a clock at position $\mathbf{r}$ with velocity $\mathbf{v}$, the proper time evolution is
\begin{equation}
	\frac{d\tau}{dt} \approx 1 - \frac{1}{c^2}\left( {W(\mathbf{r}) + \frac{1}{2}{\mathbf{v}^2}} \right) + O(c^{ - 4}).
\end{equation}
Assuming the ground clock emits a signal with a proper frequency $f_g$, the frequency received by the satellite clock is $f_s$. According to the frequency transfer equation, the frequency ratio between the satellite and the ground station can be expressed as
\begin{equation}
	\label{equ:radiooffrequency}
	\frac{f_s}{f_g} = \frac{d{\tau _g}}{d{t_g}} \left( {\frac{d{\tau _s}}{d{t_s}}} \right)^{ - 1} \frac{d{t_g}}{d{t_s}}.
\end{equation}
The first two terms represent the relativistic clock shift (gravitational redshift and second-order Doppler effect). The last term   encapsulates link effects such as the first-order Doppler shift, Shapiro delay, and atmospheric delays. Since the first-order Doppler shift is typically the dominant term in high-precision experiments, it must be canceled using advanced Doppler compensation techniques, which will be detailed in Section \ref{dopplercompensationmodel}. Once the first-order Doppler and high-precision link effects are modeled and removed, the residual fractional frequency difference is dominated by relativistic effects
\begin{equation}
	\frac{\Delta f}{f} = \frac{{f_s} - {f_g}}{f_g} \approx \frac{W(\mathbf{r}_s) - W(\mathbf{r}_g)}{c^2} + \frac{{\mathbf{v}_s}^2 - {\mathbf{v}_g}^2}{2c^2} + \varepsilon.
\end{equation}
The first term is gravitational redshift term which contains the static Earth potential and the time-varying tidal potential induced by celestial bodies. The second-order Doppler term depends strictly on the velocities of the station and satellite. The $\varepsilon$ term incorporates measurement noise, including clock instability and model errors.

\subsection{Doppler Cancellation Model}
\label{dopplercompensationmodel}
\begin{figure}
	\includegraphics[width=0.5\textwidth]{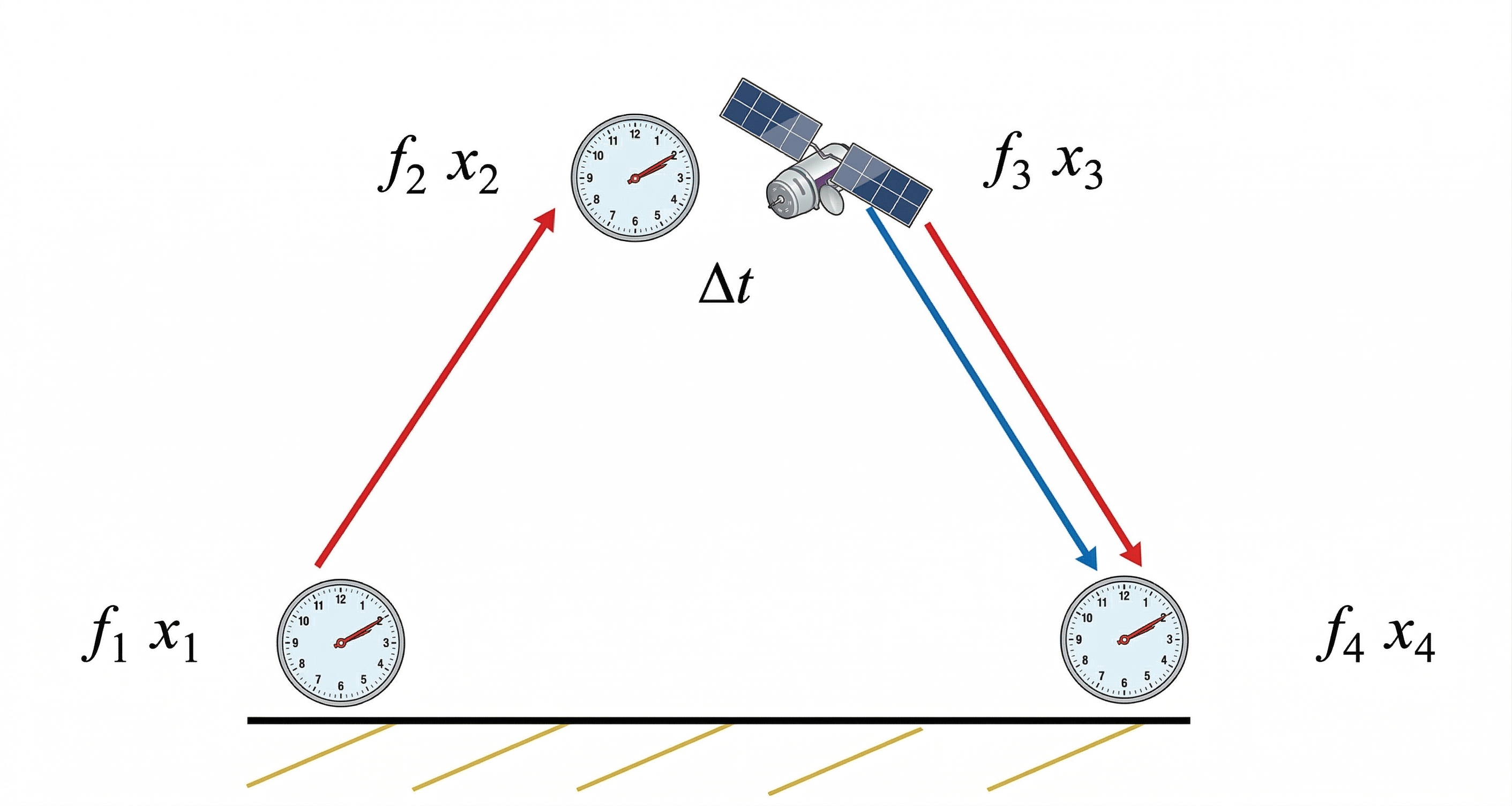}
	\caption{\label{fig:doppler} Diagram of Doppler cancellation cnfiguration. There are three frequency transfer links in the configuration: an uplink signal and two downlink signals.}
\end{figure}
In space-to-ground clock comparisons, the vast velocity difference between the satellite and the ground station generates a massive Doppler shift, which acts as the dominant noise source requiring stringent cancellation. To extract the weak relativistic and tidal signals, we employ a three-link Doppler cancellation configuration as illustrated in Figure \ref{fig:doppler}. In this configuration, the frequency transfer system comprises one uplink and two distinct downlinks, which physically constitute a coherent two-way link and an independent one-way downlink.

Specifically, as indicated by the red arrows in Figure \ref{fig:doppler}, the coherent two-way link begins with an optical signal emitted from the ground station at coordinate time $t_1$ with a proper emission frequency $f_1$. After free-space propagation, the signal is received by the satellite at coordinate time $t_2$ with a reception frequency $f_2$. Following a brief internal hardware delay $\Delta t$, the satellite employs a phase-locked transponder to coherently retransmit the signal, enforcing the assumption that the retransmitted frequency equals the received frequency ($f_3/f_2 = 1$). This reflected signal is ultimately received back at the ground station at coordinate time $t_4$ with a reception frequency $f_4$. Concurrently, as indicated by the blue arrow, the satellite's local atomic clock actively emits an independent one-way downlink signal at $t_3$ with proper frequency $f_3$, which is simultaneously received at the ground station at $t_4$.

The fractional frequency ratio of the uplink $f_2/f_1$ is rigorously evaluated by substituting the spacetime metric into the frequency transfer equation (Eq. \ref{equ:radiooffrequency}) and performing a Taylor expansion up to $\mathcal{O}(c^{-3})$ \cite{qinRelativisticFormulationDual2026}:
\begin{strip}
  \begin{equation}
    \label{equ:uplink}
    \begin{aligned}
      \frac{f_2}{f_1} &= \frac{d\tau_1}{dt_1}{\left( \frac{d\tau_2}{dt_2} \right)^{-1}}\frac{dt_1}{dt_2} \\
    	&\approx 1 - \frac{\mathbf{N}_{gs,12} \cdot \mathbf{v}_{gs,12}}{c} - \frac{\left( \mathbf{N}_{gs,12} \cdot \mathbf{v}_{gs,12} \right)\left[ \mathbf{N}_{gs,12} \cdot \mathbf{v}_g\left( t_1 \right) \right]}{c^2} + \frac{1}{c^2}\left[ W_s\left( t_2 \right) - W_g\left( t_1 \right) + \frac{{v_s}^2\left( t_2 \right)}{2} - \frac{{v_g}^2\left( t_1 \right)}{2} \right] \\
    	&\quad - \frac{ \mathbf{N}_{gs,12} \cdot \mathbf{v}_{gs,12} }{c^3}\left\{ \left[ \mathbf{N}_{gs,12} \cdot \mathbf{v}_g\left( t_1 \right) \right]^2 + W_s\left( t_2 \right) - W_g\left( t_1 \right) + \frac{{v_s}^2\left( t_2 \right)}{2} - \frac{{v_g}^2\left( t_1 \right)}{2} \right\} \\
    	&\quad - \frac{d\Delta t_{other}}{dt_2}\left\{ 1 + \frac{\mathbf{N}_{gs,12} \cdot \mathbf{v}_g\left( t_1 \right)}{c} \right\} + O\left( c^{-4} \right),
    \end{aligned}
  \end{equation}
\end{strip}
\noindent where $\mathbf{N}_{gs,12}$ is the unit vector pointing from the ground station to the satellite in uplink, $\mathbf{v}_{gs,12} = \mathbf{v}_{s}\left( t_2 \right) - \mathbf{v}_{g}\left( t_1 \right)$ is the velocity vector of the satellite in $t_2$ relative to that of the ground station in $t_1$, $W_s\left( t_2 \right)$ is the potential of the satellite at time $t_2$, $W_g\left( t_1 \right)$ is the potential of the ground station at time $t_1$, $\mathbf{v}_s\left( t_2 \right)$ is the velocity of the satellite at time $t_2$, $\mathbf{v}_g\left( t_1 \right)$ is the velocity of the ground station at time $t_1$, $\Delta t_{other}$ is the time delay caused by other sources such as atmospheric delay and gravitational delay. 

Similarly, the fractional frequency ratio for the one-way downlink $f_4/f_3$ can be rigorously derived following the same relativistic framework. Its mathematical expression exhibits a strict structural symmetry with the uplink equation, where the kinematic and potential variables are systematically interchanged to reflect the reversed satellite-to-ground propagation geometry. Specifically, the state vectors evaluated at epochs $t_1$ and $t_2$ are mapped to $t_3$ and $t_4$, respectively, and the line-of-sight unit vector is reversed from $\mathbf{N}_{gs}$ to $\mathbf{N}_{sg}$.

For the coherent two-way link, the total frequency ratio $f_4/f_1$ is the product of the uplink and downlink ratios. Because the first-order Doppler shift is roughly symmetric in the two-way transit but asymmetric in the pure relativistic terms, we can isolate the desired relativistic signature by constructing a linear combination of the observables at the ground station. The Doppler cancellation observable is defined as \cite{blanchetRelativisticTheoryTime2001}:
\begin{equation}
	\label{equ:dopplercancellation}
	\Delta  = \frac{1}{2}\frac{{{f_4}}}{{{f_1}}} - \frac{{{f_4}}}{{{f_3}}}.
\end{equation}
By substituting the expanded expressions of the uplink and downlink ratios, the dominant $\mathcal{O}(c^{-1})$ first-order Doppler shifts are exactly algebraically canceled. Retaining the kinematic and potential terms up to $\mathcal{O}(c^{-3})$, the residual fractional frequency signal simplifies to:
\begin{equation}
	\Delta  \approx  - \frac{1}{2} + {\Delta _2} + {\Delta _{signal}} + {\Delta _3} + {\Delta _L},
\end{equation}
where $-1/2$ is DC component, $\Delta _2$ is the residual second-order term, $\Delta _{signal}$ is the clock signal term, $\Delta _3$ is the residual third-order term, and $\Delta _L$ is the residual link effect term including atmospheric delay and gravitational delay. Considering that in practical ground-based clock comparisons, experimental data and measured quantities are typically recorded at the epoch of optical signal reception, the expressions must be unified in terms of the parameters evaluated at time $t_4$. After calculation, the expression for the residual second-order term can be written as
\begin{equation}
  \label{equ:residualsecondorder}
  \begin{aligned}
    \Delta _2 &= \frac{v_{sg}^2(t_4)}{2cD_{sg}(t_4)}\Delta t + \frac{\mathbf{N}_{sg}(t_4) \cdot \mathbf{a}_{sg}(t_4)}{2c}\Delta t\\
    &\quad - \frac{[\mathbf{N}_{sg}(t_4) \cdot \mathbf{v}_{sg}(t_4)]^2}{2cD_{sg}(t_4)}\Delta t + \frac{\mathbf{v}_{sg}(t_4) \cdot \mathbf{v}_g(t_4)}{c^2}\\
    &\quad + \frac{D_{sg}(t_4)[\mathbf{N}_{sg}(t_4) \cdot \mathbf{a}_g(t_4)]}{c^2}
  \end{aligned}
\end{equation}
where $\mathbf{v}_{sg}(t_4) = \mathbf{v}_{g}(t_4) - \mathbf{v}_{s}(t_4)$ and $\mathbf{a}_{sg}(t_4) = \mathbf{a}_g(t_4) - \mathbf{a}_s(t_4)$ denote the relative velocity and acceleration vectors between the ground station and the satellite at epoch $t_4$, respectively. $\mathbf{N}_{sg}(t_4)$ is the unit vector pointing from the satellite to the ground station at $t_4$, $D_{sg}(t_4)$ is the satellite-ground distance at $t_4$, and $\Delta t$ is the signal reflection delay in satellite. The expression for the clock signal term can be written as
\begin{equation}
	\label{equ:clocksignal}
  \begin{aligned}
    \Delta_{signal} &= \frac{1}{c^2}\\
    & \left[ W_s(t_4) - W_g(t_4) + \frac{v_s^2(t_4)}{2} - \frac{v_g^2(t_4)}{2} \right],
  \end{aligned}
\end{equation}
the expression for the residual third-order term can be written as
\begin{equation}
  \begin{aligned}
    \Delta_3
    &= \frac{A}{c^3}
    + \frac{B\,\Delta t}{2c^2 D_{sg}(t_4)},
  \end{aligned}
\end{equation}
with
\begin{equation*}
  \begin{aligned}
    A &= - \left[ \mathbf{N}_{sg}(t_4) \cdot \mathbf{v}_{sg}(t_4) \right]\left[ \mathbf{v}_{sg}(t_4) \cdot \mathbf{v}_g(t_4) \right]\\
    &\quad - D_{sg}(t_4)\left[ \mathbf{N}_{sg}(t_4) \cdot \mathbf{v}_{sg}(t_4) \right]\left[ \mathbf{N}_{sg}(t_4) \cdot \mathbf{a}_g(t_4) \right] \\
    &\quad - 2D_{sg}(t_4)\left[ \mathbf{v}_{sg}(t_4) \cdot \mathbf{a}_g(t_4) \right] + D_{sg}(t_4)\left[ \mathbf{v}_{sg}(t_4) \cdot \mathbf{a}_s(t_4) \right] \\
    &\quad - {D_{sg}}^2(t_4)\left[ \mathbf{N}_{sg}(t_4) \cdot \mathbf{b}_g(t_4) \right] + D_{sg}(t_4)\left[ \dot{W}_s(t_4) - \dot{W}_g(t_4) \right] \\
    &\quad + \left[ \mathbf{N}_{sg}(t_4) \cdot \mathbf{v}_{sg}(t_4) \right]\left[ W_g(t_4) - W_s(t_4) + \frac{v_g^2(t_4)}{2} - \frac{v_s^2(t_4)}{2} \right],\\
    B &= 2{v_{sg}}^2(t_4)\left[ \mathbf{N}_{sg}(t_4) \cdot \mathbf{v}_{sg}(t_4) \right] - 2\left[ \mathbf{N}_{sg}(t_4) \cdot \mathbf{v}_{sg}(t_4) \right]^3 \\
    &\quad - D_{sg}(t_4)\left[ \mathbf{v}_{sg}(t_4) \cdot \mathbf{a}_g(t_4) \right] - 4D_{sg}(t_4)\left[ \mathbf{v}_{sg}(t_4) \cdot \mathbf{a}_{sg}(t_4) \right] \\
    &\quad + 2D_{sg}(t_4)\left[ \mathbf{N}_{sg}(t_4) \cdot \mathbf{v}_{sg}(t_4) \right]\left[ \mathbf{N}_{sg}(t_4) \cdot \mathbf{a}_{sg}(t_4) \right] \\
    &\quad - 2{D_{sg}}^2(t_4)\left[ \mathbf{N}_{sg}(t_4) \cdot \mathbf{b}_g(t_4) \right] + {D_{sg}}^2(t_4)\left[ \mathbf{N}_{sg}(t_4) \cdot \mathbf{b}_s(t_4) \right] \\
    &\quad + D_{sg}(t_4)\left[ \dot{W}_s(t_4) - \dot{W}_g(t_4) \right],
  \end{aligned}
\end{equation*}
where $\mathbf{b}_g(t_4)$ and $\mathbf{b}_s(t_4)$ are the jerk of the ground station and the satellite at epoch $t_4$, and the expression for the residual link effect term can be written as
\begin{equation}
  \begin{aligned}
    \Delta_L &= \frac{d\Delta t_{other}}{dt_4}\\
    &\left\{ 2\frac{\mathbf{N}_{sg}(t_4) \cdot \mathbf{v}_g(t_4)}{c} - \frac{\mathbf{N}_{sg}(t_4) \cdot \mathbf{v}_s(t_4)}{c} \right\},
  \end{aligned}
\end{equation}
where $\Delta t_{other}$ represents the time delay introduced by the signal link.

\subsection{Earth Gravitational and Tidal Potentials}
\label{potentialmodel}
Due to the heterogeneous mass distribution and non-spherical shape of the Earth, its gravitational potential is conventionally expanded using Spherical Harmonics
\begin{equation}
  \begin{aligned}
	  W_E(r,\phi,\lambda) &= \frac{GM_E}{r} \sum_{n=0}^N \sum_{m=0}^n \left(\frac{R_E}{r}\right)^n \bar{P}_{nm}(\sin\phi) \\
    &\quad \left[ \bar{C}_{nm} \cos(m\lambda) + \bar{S}_{nm} \sin(m\lambda) \right]
  \end{aligned}
\end{equation}
where $G{M_E}$ is the Earth's standard gravitational parameter, $R_E$ is the equatorial radius, ${\bar P_{nm}}\left( {\sin \phi } \right)$ are the fully normalized associated Legendre polynomials, and ${\bar C_{nm}},{\bar S_{nm}}$ are the normalized spherical harmonic coefficients derived from satellite gravity missions (e.g., GRACE, GOCE). In high-precision clock comparison calculations, the accuracy of the gravitational potential affects the accuracy of frequency shift calculations. For a satellite, due to the rapid spatial attenuation with distance $r$, truncating the model to tens of degrees is sufficient to achieve $10^{-18}$ precision. Conversely, computing the precise local potential at the ground station requires significantly higher-degree models combined with local gravimetric measurements.

In celestial mechanics, the tidal potential is the differential gravitational potential exerted by third bodies (mainly the Moon and the Sun) after removing the inertial term due to the geocenter's translation. This potential is the fundamental source of solid Earth tides, ocean tides, and satellite orbital perturbations. For a celestial body of mass $M_p$ at a geocentric position $\mathbf{r}_p$, the tidal potential at spatial position $\mathbf{r}$ is
\begin{equation}
	{W_{tide}}(\mathbf{r}) = \sum_p GM_p\left( \frac{1}{|\mathbf{r}_p - \mathbf{r}|} - \frac{1}{|\mathbf{r}_p|} - \frac{\mathbf{r} \cdot \mathbf{r}_p}{|\mathbf{r}_p|^3} \right)
\end{equation}
In the near-Earth space space ($r \ll {r_P}$), this can be expanded into a converging series using Legendre polynomials
\begin{equation}
	{W_{tide}}(\mathbf{r}) = \sum_p \frac{GM_p}{r_p}\sum_{n=2}^\infty \left(\frac{r}{r_p}\right)^n P_n(\cos\psi_p),
\end{equation}
where $\psi _p$ is the geocentric zenith angle between $\mathbf{r}$ and $\mathbf{r}_p$.

\subsection{Solid Earth Tide Model}
\label{solidtidemodel}
The Earth is not a perfectly rigid body. Under the tidal forces of the Moon and Sun, the solid Earth (crust and mantle) undergoes periodic elastic deformations. These deformations manifest as vertical and horizontal surface displacements, and perturbations in the Earth's inherent gravitational field. These linear geodynamic responses are described by dimensionless parameters known as Love numbers (${h_n},{k_n},{l_n}$). The displacement response is parameterized as \cite{greenJourneyTides2022}
\begin{equation}
	\begin{aligned}
		u_r &= \sum_{n=2}^\infty \frac{h_n}{g}W_n \\
		u_\theta &= \sum_{n=2}^\infty \frac{l_n}{g}\frac{\partial W_n}{\partial\theta} \\
		u_\lambda &= \sum_{n=2}^\infty \frac{l_n}{g\sin\theta}\frac{\partial W_n}{\partial\lambda},
	\end{aligned}
\end{equation}
where $W_n$ is the degree-$n$ tidal potential, $g$ is the local surface gravity, $\theta$ and $\lambda$ are the latitude and longitude of the Earth. Following IERS conventions, the degree-2 correction is the most significant. The degree-2 displacement vector caused by a celestial body is
\begin{equation}
  \label{equ:degree2displacement}
  \begin{aligned}
    \delta\mathbf{r}_{E,2} &= \sum_p \frac{GM_p R_E^4}{GM_E R_p^3} \Bigg\{h_2 \hat{\mathbf{r}}\Bigg(\frac{3(\hat{\mathbf{R}}_p\cdot\hat{\mathbf{r}})^2-1}{2}\Bigg) \\
    &\quad + 3l_2(\hat{\mathbf{R}}_p\cdot\hat{\mathbf{r}})\left[\hat{\mathbf{R}}_p- (\hat{\mathbf{R}}_p\cdot\hat{\mathbf{r}})\hat{\mathbf{r}}\right]\Bigg\},
  \end{aligned}
\end{equation}
where $GM_p$ is the gravitational constant of the celestial body, $GM_E$ is the gravitational constant of the Earth, $R_E$ is the mean equatorial radius of the Earth, $R_p$ is the distance from the center of the Earth to the celestial body, $\hat{\mathbf{r}}$ is the unit vector representing the mean geocentric position of the station, and $\hat{\mathbf{R}}_p$ is the unit vector pointing from the center of the Earth toward the celestial body. Similarly, the third-order displacement vector is
\begin{equation}
  \begin{aligned}
  \delta\mathbf{r}_{E,3} &= \sum_p \frac{G M_p R_E^5}{G M_E R_p^4} \Bigg\{h_3 \hat{\mathbf{r}}\left[\frac{5}{2}\left(\hat{\mathbf{R}}_p \cdot \hat{\mathbf{r}}\right)^3 - \frac{3}{2}\left(\hat{\mathbf{R}}_p \cdot \hat{\mathbf{r}}\right)\right] \\
  &\quad + l_3\left[\frac{15}{2}\left(\hat{\mathbf{R}}_p \cdot \hat{\mathbf{r}}\right)^2 - \frac{3}{2}\right]\left[\hat{\mathbf{R}}_p - \left(\hat{\mathbf{R}}_p \cdot \hat{\mathbf{r}}\right)\hat{\mathbf{r}}\right]\Bigg\}.
  \end{aligned}
\end{equation}
While degree-2 radial displacements can reach tens of centimeters, degree-3 displacements are merely on the millimeter scale and only require consideration under extreme precision demands. The deformation also redistributes the Earth's mass, generating an additional geopotential increment parameterized by the Love number $k_n$
\begin{equation}
	\label{equ:additionalpotential}
	\Delta W\left( r \right) = \sum\limits_{n = 2}^\infty  {{k_n}{{\left( {\frac{{{R_E}}}{r}} \right)}^{n + 1}}{W_n}\left( {{R_E}} \right)}.
\end{equation}

\section{Noise Model}
\label{noisemodel}
\subsection{Clock Noise}
\label{clocknoise}
Atomic clocks serve as frequency reference sources, but their stability is fundamentally limited by internal quantum physics, electronic thermal noise, and environmental perturbations. Clock stability is conventionally quantified using the Allan variance in the time domain, and the Power Spectral Density (PSD) in the frequency domain. The clock's output signal is typically modeled as \cite{barnesCharacterizationFrequencyStability1971}
\begin{equation}
	V(t) = [V_0 + \varepsilon(t)]\sin[2\pi\nu_0 t + \delta\phi(t)],
\end{equation}
where $V_0$ is the nominal amplitude, $\nu _0$ is the nominal frequency, and $\varepsilon \left( t \right),\delta \phi \left( t \right)$ represents the amplitude and phase fluctuations. The phase deviation of an atomic clock is
\begin{equation}
	x(t) = \frac{\delta\phi(t)}{2\pi\nu_0},
\end{equation}
which represents the deviation between the clock reading and the ideal time, the relative frequency deviation can be written as
\begin{equation}
	y(t) = \frac{dx(t)}{dt} = \frac{1}{2\pi\nu_0}\frac{d}{dt}\delta\phi(t).
\end{equation}
The PSD of frequency fluctuations $y\left( t \right)$ in precision oscillators can typically be well approximated as a linear combination of a series of power-law functions \cite{allanStatisticsAtomicFrequency1966}
\begin{equation}
	S_y(f) = \sum_{\alpha=-2}^2 h_\alpha f^\alpha.
\end{equation}
Based on the exponent $\alpha$, clock noise is categorized into five basic types in Table \ref{tab:clocknoise}. For long-baseline periodic tidal signals spanning thousands of seconds, White Phase Noise (WPM) acts mainly as the measurement system's noise floor and is rapidly averaged out (Allan deviation scales as ${\tau ^{ - 1}}$). Therefore, we synthesize the clock noise strictly through a combination of White Frequency Noise (WFM) and Flicker Frequency Noise (FFM).
\begin{table}[t]
  \centering
  \caption{Clock Noise Characteristics.}
  \label{tab:clocknoise}
  \begin{tabular}{lccc}
    \toprule
    Noise Type & PSD (${S_y}\left( f \right)$) & ADEV ($\sigma_y\left( \tau \right)$) \\
    \midrule
    WPM & $\propto f^2$ & $\propto \tau^{-1}$ \\
    FPM & $\propto f^1$ & $\propto \tau^{-1}$ \\
    WFM & $\propto f^0$ & $\propto \tau^{-1/2}$ \\
    FFM & $\propto f^{-1}$ & $\propto \tau^0$ \\
    RWFM & $\propto f^{-2}$ & $\propto \tau^{1/2}$ \\
    \bottomrule
  \end{tabular}
\end{table}

For power-law noise, the Kasdin algorithm can be used to quickly generate time-domain sequences that match the noise power spectrum. For each time step $n$, the colored noise sequence can be expressed as a convolution of a white noise sequence \cite{kasdinDiscreteSimulationColored1995}
\begin{equation}
	{x_n} = \sum\limits_{k = 0}^n {{h_k}{w_{n - k}}},
\end{equation}
where $x_n$ is a Gaussian white noise sequence with mean zero and unit variance, and the impulse response coefficients $h_k$ are determined by the recursive relation
\begin{equation}
	\begin{aligned}
		{h_0} &= 1\\
		h_k &= \left(\frac{\alpha}{2} + k - 1\right)\frac{h_{k-1}}{k}, \quad k = 1,2,\ldots
	\end{aligned}
\end{equation}
where $\alpha$ is the power of the noise power spectrum.

\subsection{Satellite Precise Orbit Determination Noise}
\label{satellitenoise}
POD errors primarily stem from imperfect dynamic models and observational geometry limitations. For the IGSO satellite selected in this study (altitude ~36,000 km), atmospheric drag is negligible, making Solar Radiation Pressure (SRP) the dominant non-conservative perturbation \cite{xiaAdvancingSolarRadiation2022,qinPreciseOrbitDetermination2019,zhaoAnalysisLongtermDynamical2015}. Current analytical SRP models (e.g., ECOM, Box-Wing) fail to absorb all non-gravitational accelerations completely, leaving residuals that accumulate into position drift via orbit integration \cite{rodriguez-solanoAdjustableBoxwingModel2012,duanImprovingSolarRadiation2020}. Furthermore, high-degree truncation errors of the gravity field, third-body ephemeris errors, and Earth Orientation Parameter (EOP) prediction biases all contribute to POD inaccuracies.

Orbit errors exhibit strong anisotropy in the Radial-Transverse-Normal (RTN) frame. Radial error directly couples into the gravitational potential computation but is typically well-constrained by ground station geometry (usually yielding better precision than the transverse/along-track axes). The along-track error is heavily coupled with orbital energy errors governed by Kepler's third law, usually exhibiting the largest amplitude and a distinct linear drift trend \cite{akiyamaAlongtrackOrbitError2025,yuCovarianceAnalysisRealTime2019}.

Critically, POD noise is not ideal Gaussian white noise. It exhibits discrete spectral structures dominated by the 1-CPR (Cycle Per Revolution) periodic oscillation caused by the propagation of initial state vector errors in the central force field. For IGSO satellites, this 1-CPR frequency closely aligns with the diurnal diurnal tidal components (e.g., $O_1,K_1$), severely risking signal aliasing during parameter estimation \cite{duanImprovingSolarRadiation2020,yuCovarianceAnalysisRealTime2019,staceyPROCESSNOISECOVARIANCE2021,duanEnhancedSolarRadiation2021,leonardGravityErrorCompensation2013}. We model the orbital error as a mixed signal comprising deterministic harmonics (1-CPR, 2-CPR), linear drift, and a first-order Gauss-Markov process to accurately reflect its spectral traits. We decompose the orbital position error vector of the satellite $\Delta \mathbf{r}(t)$ into three components in the RTN coordinate system and construct a time-series generation model
\begin{equation}
  \begin{aligned}
    {x_k}(t) &= {B_k} + {\beta _k}t\\
    & + \sum_{j=1}^2 A_{k,j} \cos(j\cdot\omega_{orb}t + \varphi_{k,j}) + \varepsilon_k(t), \\
    &k \in \{R, T, N\},
  \end{aligned}
\end{equation}
where $B_k$ is used to simulate errors in the radial fixation system, and $\beta _k$ is the linear drift coefficient, primarily used to simulate long-term positional drift in the tangential direction caused by energy dissipation or errors in the semi-major axis estimation. Given that real-time satellite precise orbit determination corrects for long-term linear drift terms, this term is set to 0 in the actual simulation. $A_{k,j}$ and $\varphi _{k,j}$ denote the amplitude and phase of the th harmonic. ${\omega _{orb}} = 2\pi /{T_{orb}}$ is the satellite orbital frequency, and the model includes the fundamental frequency (1-CPR) and the second harmonic (2-CPR) terms, which correspond to the error propagation of the initial state vector and the higher-order residuals of the solar radiation pressure model, respectively. ${\varepsilon _k}\left( t \right)$ represents the residual colored noise, generated using a first-order Gauss-Markov process
\begin{equation}
	{\varepsilon _k}({t_{i + 1}}) = {e^{ - \Delta t/{\tau _k}}}{\varepsilon _k}({t_i}) + {\sigma _{noise,k}}\sqrt {1 - {e^{ - 2\Delta t/{\tau _k}}}} {w_i}
\end{equation}
where $\tau _k$ is the relevant time, reflecting the temporal memory of the error, $\sigma _{noise,k}$ is the standard deviation of the drive noise, and $w_i$ is a white noise sequence following a standard normal distribution.

\section{Simulation and Parameter Estimation}
\label{simulation}
\subsection{Simulation Configuration}
\label{experimentconfiguration}
To evaluate the capability of clock comparisons to invert solid Earth tide parameters, we developed a space-to-ground clock comparison simulation. The simulation is conducted within the GCRS, which serves as the inertial frame, while the Earth's gravitational field and the ground station coordinates are defined in the ITRS. Coordinate transformations between these reference systems strictly adhere to the IERS 2010 conventions.

An IGSO is selected for the satellite. This orbit geometry not only guarantees extended common-view durations with the ground station, but its high altitude also ensures that the satellite is significantly exposed to lunisolar tidal perturbations. The initial epoch is set to 00:00:00 UTC on July 1, 2007, with the initial Keplerian elements (in the GCRS) specified in Table \ref{tab:initialorbit}.
\begin{table}[t]
	\centering
	\caption{\label{tab:initialorbit}Initial orbital elements of the satellite.}
	\begin{tabular}{lcc}
		\toprule
		Parameter & Value & Unit \\
		\midrule
		Semi-major axis ($a$) & 42164.2 & km \\
		Eccentricity ($e$) & 0.007 & degree \\
		Inclination ($i$) & 20 & degree \\
		RAAN ($\Omega$) & 310 & degree \\
		Argument of periapsis ($\omega$) & 90 & degree \\
		Mean Anomaly ($M$) & 0 & degree \\
		\bottomrule
	\end{tabular}
\end{table}

\begin{figure}
	\includegraphics[width=0.5\textwidth]{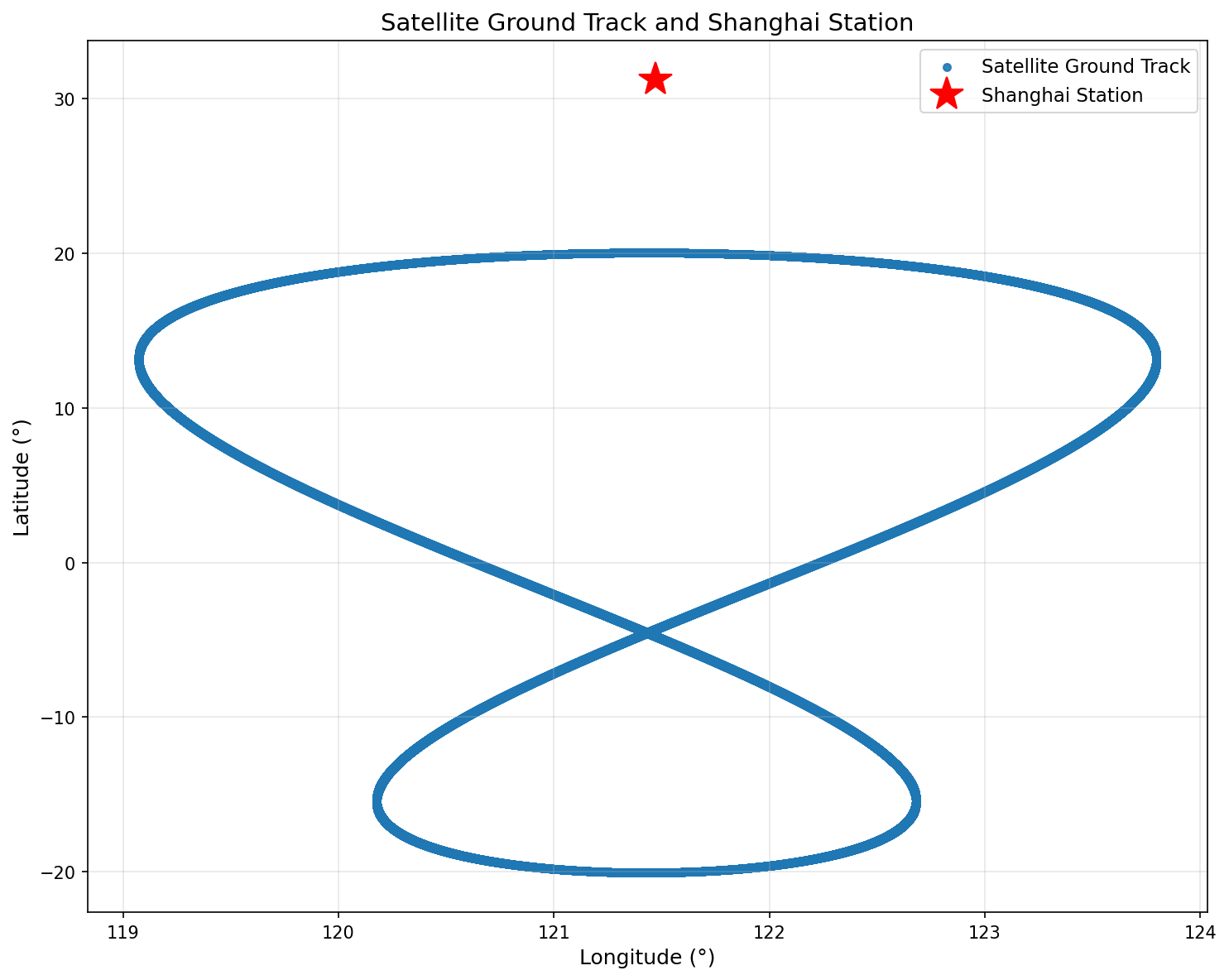}
	\caption{\label{fig:orbit} Single-period satellite ground track and the geographic location of the Shanghai ground station in the ITRS. The blue solid line illustrates the characteristic figure-eight ground track of the IGSO satellite, while the red star indicates the position of the Shanghai station.}
\end{figure}

The satellite trajectory is propagated using a high-precision Dormand-Prince 8th-order numerical integrator. The Earth's background static gravity field is modeled using the EGM2008 model \cite{pavlisDevelopmentEvaluationEarth2012}, with the spherical harmonic coefficients expanded up to degree and order $200 \times 200$. The Sun and the Moon are incorporated as the primary tidal perturbing bodies. Solar radiation pressure (which is separately introduced as an orbital noise component) and atmospheric drag (which is negligible for such high altitudes) are intentionally omitted from the baseline dynamical propagation. The Shanghai Astronomical Observatory is designated as the ground station, with its mean position in the ITRS denoted as (121.47°E, 31.23°N, 5 m). To support the feasibility of the parameter estimation, it is assumed that the three-dimensional position of the ground station can be independently determined to the centimeter level using geophysical measurement methods \cite{altamimiITRF2014NewRelease2016,zhangAssessmentImprovementObservation2024}. The single-period satellite ground track and the geographic location of the ground station are illustrated in Figure \ref{fig:orbit}.

\begin{figure}
	\includegraphics[width=0.5\textwidth]{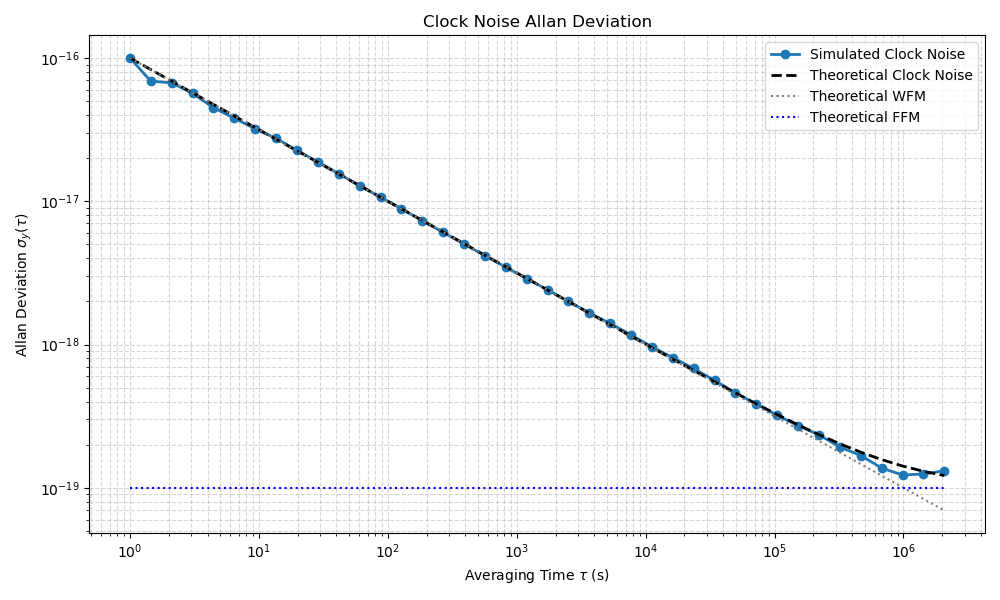}
	\caption{\label{fig:clocknoise} ADEV of the generated clock noise. The blue curve represents the simulated clock stability, which follows the theoretical WFM noise slope of $\tau^{-1/2}$ at shorter averaging times and gradually approaches a FFM floor of $10^{-19}$ at longer averaging times}
\end{figure}

\begin{figure}
	\includegraphics[width=0.5\textwidth]{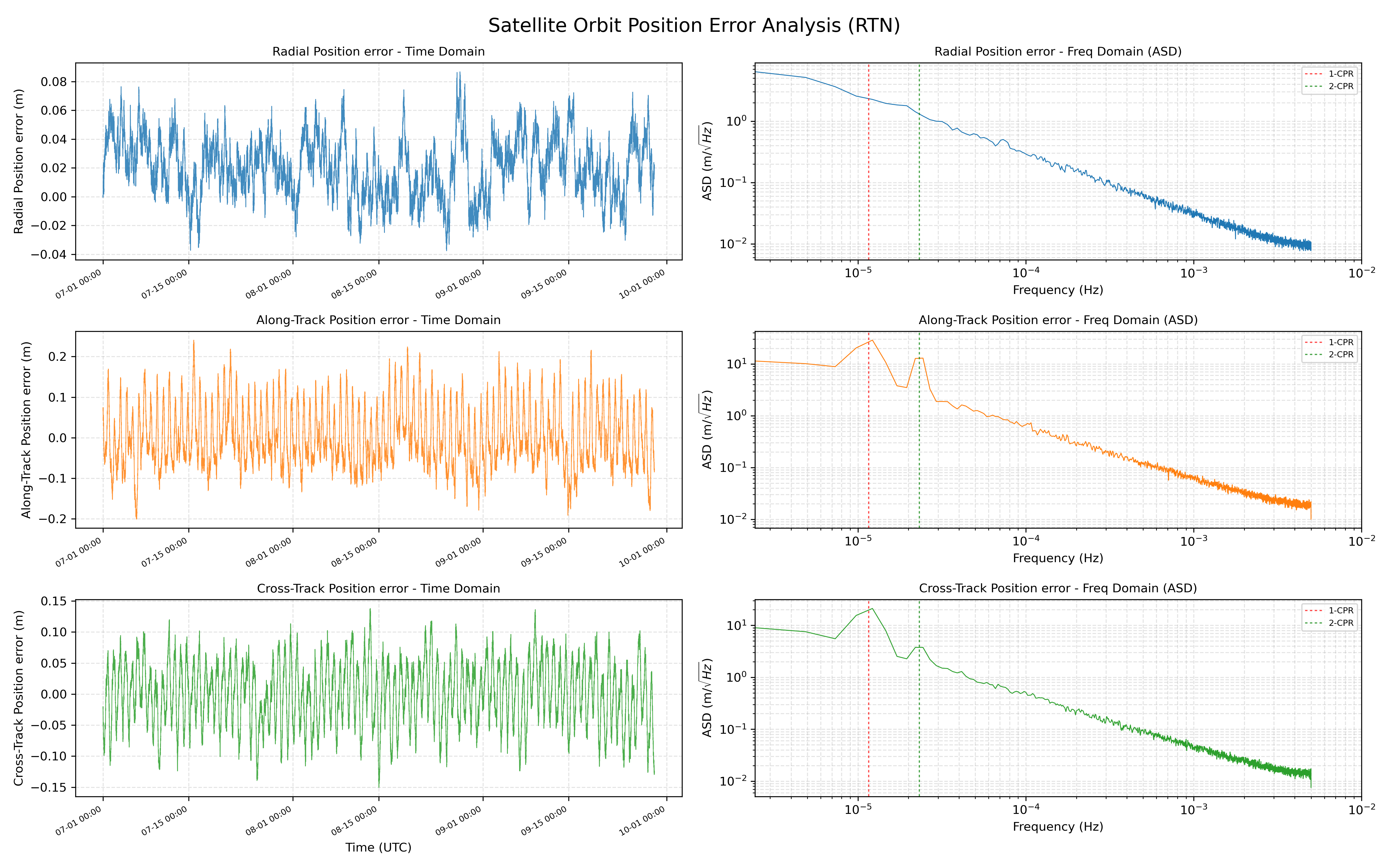}
	\caption{\label{fig:orbitnoise} Time-domain and frequency-domain characteristics of the injected satellite POD errors.}
\end{figure}

To emulate real operational constraints, we systematically inject the synthesized noise models (defined in Section \ref{noisemodel}) into the ideal theoretical observables. For clock noise, we inject both WFM and FFM into the frequency comparison link. The WFM limits the short-term high-frequency sampling precision, scaling as $\tau^{-1/2}$, while the FFM establishes the ultimate Allan Deviation (ADEV) floor for long integration times, scaling as $\tau^{0}$. In our simulation, the clock noise floor is conservatively anchored at an Allan deviation of $1.0 \times 10^{-19}$ and the parameter of WFM is set to $1 \times {10^{ - 16}}\sqrt \tau  $ as shown by Figure \ref{fig:clocknoise}, representing the state-of-the-art capability of current optical lattice clocks \cite{liuZeroDeadTimeStrontiumLattice2025}. The satellite POD noise is synthesized using the RTN mixed-model framework. The radial error (R), which is the most critical axis as it directly couples into the geopotential calculation, is assigned an RMS precision of 3-5 cm. The along-track (T) and cross-track (N) errors, governed heavily by orbital energy uncertainties and orbital plane orientation respectively, are assigned RMS precisions of 8-12 cm and 8-10 cm \cite{liApplicationAccuracyAnalysis2025}. These noise series incorporate the 1-CPR and 2-CPR sinusoidal oscillations to mimic the dynamical propagation of state vector errors and Solar Radiation Pressure (SRP) mismodeling, superimposed with the first-order Gauss-Markov temporal correlation (correlation time $\tau_c = 43200$ s). The time series and ASD of the generated POD noise are shown in Figure \ref{fig:orbitnoise}.

\subsection{Signal Extraction and Parameter Estimation}
\label{signalextractionandinversion}
During the clock comparison process, the influence of solid Earth tides manifests primarily as perturbations to both the geocentric position of the ground station and the Earth's background gravitational field. To extract the desired tidal signature in our experimental setup, we decompose the residual terms obtained after Doppler cancellation. By explicitly incorporating the tidal-induced station displacement $\delta {\mathbf{r}_g},\delta {\mathbf{v}_g}$ and satellite POD errors $\Delta {\mathbf{r}_s},\Delta {\mathbf{v}_s}$ into the analytical model, the solid tide signal can be effectively isolated.

Since the time tags are aligned with the signal reception epoch at the ground station, we linearize the gravitational potentials and velocities around their nominal, unperturbed states. Let $\mathbf{r}_{s0}$ and $\mathbf{r}_{g0}$ denote the nominal geocentric position vectors of the satellite and the ground station, respectively. The true instantaneous potentials can be expanded as
\begin{equation}
	\label{equ:potentialexpand1}
	\begin{aligned}
		W_g(t_4) &= W_E(\mathbf{r}_{g0}) + \nabla W_E(\mathbf{r}_{g0}) \cdot \delta \mathbf{r}_g\\
    & + W_{tide}(\mathbf{r}_{g0}) + \nabla W_{tide}(\mathbf{r}_{g0}) \cdot \delta \mathbf{r}_g + \Delta W(\mathbf{r}_{g}) \\
		W_s(t_4) &= W_E(\mathbf{r}_{s0}) + \nabla W_E(\mathbf{r}_{s0}) \cdot \Delta \mathbf{r}_s\\
    & + W_{tide}(\mathbf{r}_{s0}) + \nabla W_{tide}(\mathbf{r}_{s0}) \cdot \Delta \mathbf{r}_s + \Delta W(\mathbf{r}_{s}),
	\end{aligned}
\end{equation}
where $W_E$ is the static Earth gravitational potential, $W_{tide}$ is the direct lunisolar tidal potential, and $\Delta W$ is the potential perturbation induced by the mass redistribution of the solid Earth (parameterized by the Love number $k_n$). Furthermore, by incorporating the definitions of tidal potential and Love numbers established in Section \ref{potentialmodel} and \ref{solidtidemodel}, Eq. \ref{equ:potentialexpand1} can be reformulated as
\begin{strip}
\begin{equation}
	\begin{aligned}
		W_g(t_4) &= W(\mathbf{r}_{g0}) + \sum_{n=2}^\infty (k_n - h_n)W_n(\mathbf{r}_{g0}) + \sum_{n=2}^\infty \left[\nabla W_n(\mathbf{r}_{g0}) - k_n(n+1)W_n(\mathbf{r}_{g0})\frac{\hat{\mathbf{r}}_g}{R_E}\right] \cdot \delta\mathbf{r}_g \\
		W_s(t_4) &= W\left( {{\mathbf{r}_{s0}}} \right) + \sum\limits_{n = 2}^\infty  {{k_n}{{\left( {\frac{{{R_E}}}{{{r_{s0}}}}} \right)}^{n + 1}}{W_n}\left( {{R_E}} \right)}  + \nabla {W_E}\left( {{\mathbf{r}_{s0}}} \right) \cdot \Delta {\mathbf{r}_s} \\
		&\quad + \sum_{n=2}^\infty \left[\nabla W_n(\mathbf{r}_{s0}) - k_n W_n(R_E)(n+1)\left(\frac{R_E}{r_{s0}}\right)^{n+1}\frac{\hat{\mathbf{r}}_{s0}}{r_{s0}}\right] \cdot \Delta\mathbf{r}_s.
	\end{aligned}
\end{equation}
\end{strip}
\noindent Similarly, the squared velocity terms can be approximated by keeping only the first-order perturbation terms
\begin{equation}
	\begin{aligned}
		{v_g}^2(t_4) &\approx {v_{g0}}^2 + 2 \mathbf{v}_{g0} \cdot \delta \mathbf{v}_g \\
		{v_s}^2(t_4) &\approx {v_{s0}}^2 + 2 \mathbf{v}_{s0} \cdot \Delta \mathbf{v}_s.
	\end{aligned}
\end{equation}
Substituting these expansions into the clock signal term (Eq. \ref{equ:clocksignal}), we can decouple the deterministic nominal signal from the solid tide perturbations and noise
\begin{equation}
	\Delta_{signal} = \Delta_{signal}^{(0)} + \Delta_{solid} + \epsilon_{signal},
\end{equation}
where $\Delta_{signal}^{(0)}$ is baseline frequency shift determined by the nominal orbits and average station location. The solid tide perturbation term $\Delta_{solid}$ is given by
\begin{equation}
	\Delta_{solid} = \frac{1}{{{c^2}}}\sum\limits_{n = 2}^\infty  {\left\{ {\left[ {{{\left( {{R_E}/{r_{s0}}} \right)}^{n + 1}} - 1} \right]{k_n} + {h_n}} \right\}{W_n}\left( {{\mathbf{r}_{g0}}} \right)},
\end{equation}
and the noise induced by satellite orbit determination errors and clock instability is isolated as
\begin{strip}
\begin{equation}
	\begin{aligned}
		\epsilon_{signal} &= \frac{1}{{{c^2}}}\left\{ {\nabla {W_E}\left( {{\mathbf{r}_{s0}}} \right) + \sum\limits_{n = 2}^\infty  {\left[ {\nabla {W_n}\left( {{\mathbf{r}_{s0}}} \right) - {k_n}{W_n}\left( {{R_E}} \right)\left( {n + 1} \right){{\left( {\frac{{{R_E}}}{{{r_{s0}}}}} \right)}^{n + 1}}\frac{{{{\hat {\mathbf{r}}}_{s0}}}}{{{r_{s0}}}}} \right]} } \right\} \cdot \Delta {\mathbf{r}_s} \\
		&\quad - \frac{1}{{{c^2}}}\sum\limits_{n = 2}^\infty  {\left[ {\nabla {W_n}\left( {{\mathbf{r}_{g0}}} \right) - {k_n}\left( {n + 1} \right){W_n}\left( {{\mathbf{r}_{g0}}} \right)\frac{{{{\hat {\mathbf{r}}}_{g0}}}}{{{R_E}}}} \right]}  \cdot \delta {\mathbf{r}_{g}} \\
		&\quad + \frac{1}{{{c^2}}}\left( {\mathbf{v}_{s0} \cdot \Delta \mathbf{v}_s - \mathbf{v}_{g0} \cdot \delta \mathbf{v}_g} \right) 
	\end{aligned}
\end{equation}
\end{strip}
\noindent Applying the same perturbation expansion to the residual second-order term (Eq. \ref{equ:residualsecondorder}). The unmodeled components driven by satellite POD errors are absorbed into a residual noise term $\epsilon_{\Delta 2}$
\begin{equation}
  \epsilon_{\Delta 2} = \frac{M\Delta t}{2c\left| \mathbf{r}_{sg} \right|} + \frac{N}{c^2},
\end{equation}
with
\begin{equation*}
  \begin{aligned}
    M &= 2\left[ \mathbf{v}_{sg} \cdot \left( \delta \mathbf{v}_g - \Delta \mathbf{v}_s \right) \right] + \left[ \left( \delta \mathbf{r}_g - \Delta \mathbf{r}_s \right) \cdot \mathbf{a}_{sg} \right] \\
			& - 2\frac{\mathbf{N}_{sg} \cdot \mathbf{v}_{sg}}{\left| \mathbf{r}_{sg} \right|}\left[ \left( \delta \mathbf{r}_g - \Delta \mathbf{r}_s \right) \cdot \mathbf{v}_{sg} \right]\\
      & - 2\left( \mathbf{N}_{sg} \cdot \mathbf{v}_{sg} \right)\left[ \mathbf{N}_{sg} \cdot \left( \delta \mathbf{v}_g - \Delta \mathbf{v}_s \right) \right] \\
			& + \frac{{3{{\left( {{{\bf{N}}_{sg}} \cdot {{\bf{v}}_{sg}}} \right)}^2}}}{{\left| {{{\bf{r}}_{sg}}} \right|}}\left[ {{{\bf{N}}_{sg}} \cdot \left( {\delta {{\bf{r}}_g} - \Delta {{\bf{r}}_s}} \right)} \right]\\
      & - \frac{{{v_{sg}}^2}}{{\left| {{{\bf{r}}_{sg}}} \right|}}\left[ {{{\bf{N}}_{sg}} \cdot \left( {\delta {{\bf{r}}_g} - \Delta {{\bf{r}}_s}} \right)} \right]\\
      & - \frac{{{{\bf{r}}_{sg}} \cdot {{\bf{a}}_{sg}}}}{{\left| {{{\bf{r}}_{sg}}} \right|}}\left[ {{{\bf{N}}_{sg}} \cdot \left( {\delta {{\bf{r}}_g} - \Delta {{\bf{r}}_s}} \right)} \right],\\
    N &=  \left( \mathbf{v}_{sg} \cdot \delta \mathbf{v}_g \right) + \left[ \left( \delta \mathbf{v}_g - \Delta \mathbf{v}_s \right) \cdot \mathbf{v}_g \right]\\
    & + \left[ \left( \delta \mathbf{r}_g - \Delta \mathbf{r}_s \right) \cdot \mathbf{a}_g \right].
  \end{aligned}
\end{equation*}
Consequently, after removing the nominal deterministic terms ($\Delta_{signal}^{(0)}$ and $\Delta_2^{(0)}$) and compensating for the link delay $\Delta_L$, we additionally subtract the nominal third-order term $\Delta_3^{(0)}$. Given that the magnitude of the third-order effect is merely on the order of $10^{-17}$, its residual variation after the nominal subtraction is entirely negligible. Thus, the final extracted fractional frequency observable can be formulated as
\begin{equation}
	y_{obs} = \Delta_{solid} + \epsilon_{total},
\end{equation}
where the total observation noise is $\epsilon_{total} = \epsilon_{signal} + \epsilon_{\Delta 2} + \epsilon_{clock}$. When restricting our estimation to the dominant degree-2 solid Earth tides ($n=2$), we substitute the theoretical displacement models (Eq. \ref{equ:degree2displacement}) and potential variations (Eq. \ref{equ:additionalpotential}) into the signal expression. The extracted degree-2 tidal signal simplifies to
\begin{equation}
	\Delta_{solid}^{(2)} = \frac{1}{{{c^2}}} {\left\{ {\left[ {{{\left( {{R_E}/{r_{s0}}} \right)}^3} - 1} \right]{k_2} + {h_2}} \right\}{W_2}\left( {{\mathbf{r}_{g0}}} \right)}.
\end{equation}
It should be noted that the Shida number $l_2$, which characterizes the tangential displacement of the ground station, is omitted from this formulation. Since the clock comparison observable is fundamentally driven by variations in the gravitational potential, the contribution of tangential displacements is negligible compared to the radial deformations and mass redistributions governed by $h_2$ and $k_2$. For high-altitude satellites, such as those in IGSO, the direct tidal potential $W_2(\mathbf{r}_{s0})$ at the satellite's altitude is significantly attenuated. This specific geometric constraint induces a strong collinearity between the sensitivity coefficients of $h_2$ and $k_2$, rendering them difficult to decouple independently in a single-satellite comparison experiment. Consequently, in practical data processing, it is often necessary to treat them as a unified effective parameter, estimating them as a specific linear combination rather than individual variables. To perform the precise parameter estimation, we construct a linearized observation equation by combining the extracted continuous time-series signal $y_{obs}$ with the astronomical ephemerides of the Moon and Sun
\begin{equation}
	Y = H X + \epsilon_{total},
\end{equation}
where $Y$ is the observation vector composed of the $y_{obs}$ time-series, and $X$ is the state vector containing the target solid Earth tide parameters. The design matrix $H$ incorporates the partial derivatives of the tidal signal $\Delta_{solid}^{(2)}$ with respect to the state parameters. To account for the colored characteristics of the orbital and clock noises, we employ a Weighted Least Squares (WLS) estimator. Defining the weight matrix $W$ as the inverse of the rigorously derived noise covariance matrix ($W = \Sigma_{\epsilon_{total}}^{-1}$) from Section \ref{noisemodel}, the objective function is formulated to minimize the weighted sum of squared residuals
\begin{equation}
	J(X) = (Y - H X)^T W (Y - H X).
\end{equation}
Minimizing this objective function yields the optimal analytical solution for the target Love numbers
\begin{equation}
	\hat{X} = (H^T W H)^{-1} H^T W Y.
\end{equation}

\subsection{Results and Discussion}
\label{result}
\begin{figure}
	\includegraphics[width=0.5\textwidth]{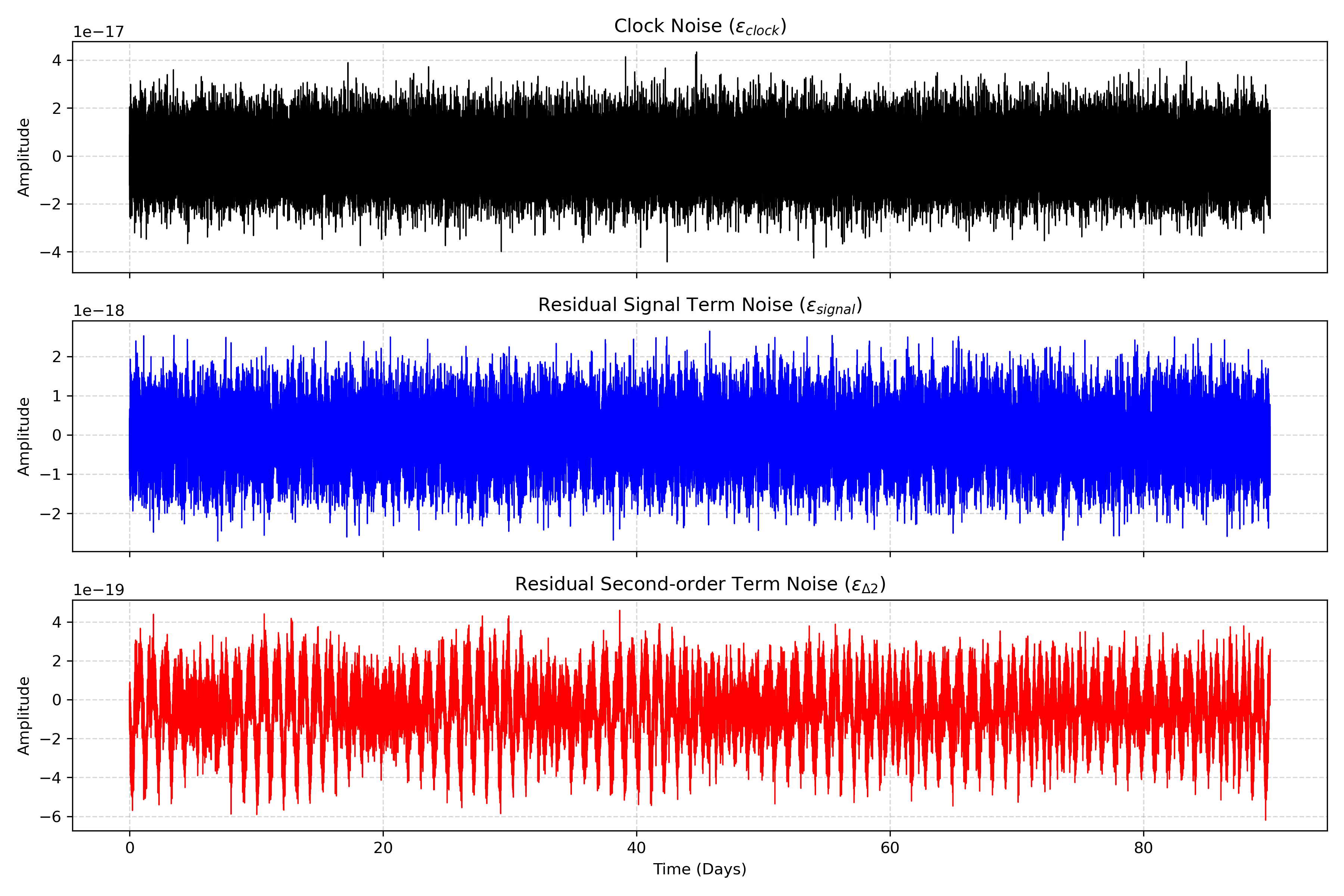}
	\caption{\label{fig:noiseseries} Time-domain series of extracted noise components over a 90-day simulation.}
\end{figure}

\begin{figure}
	\includegraphics[width=0.5\textwidth]{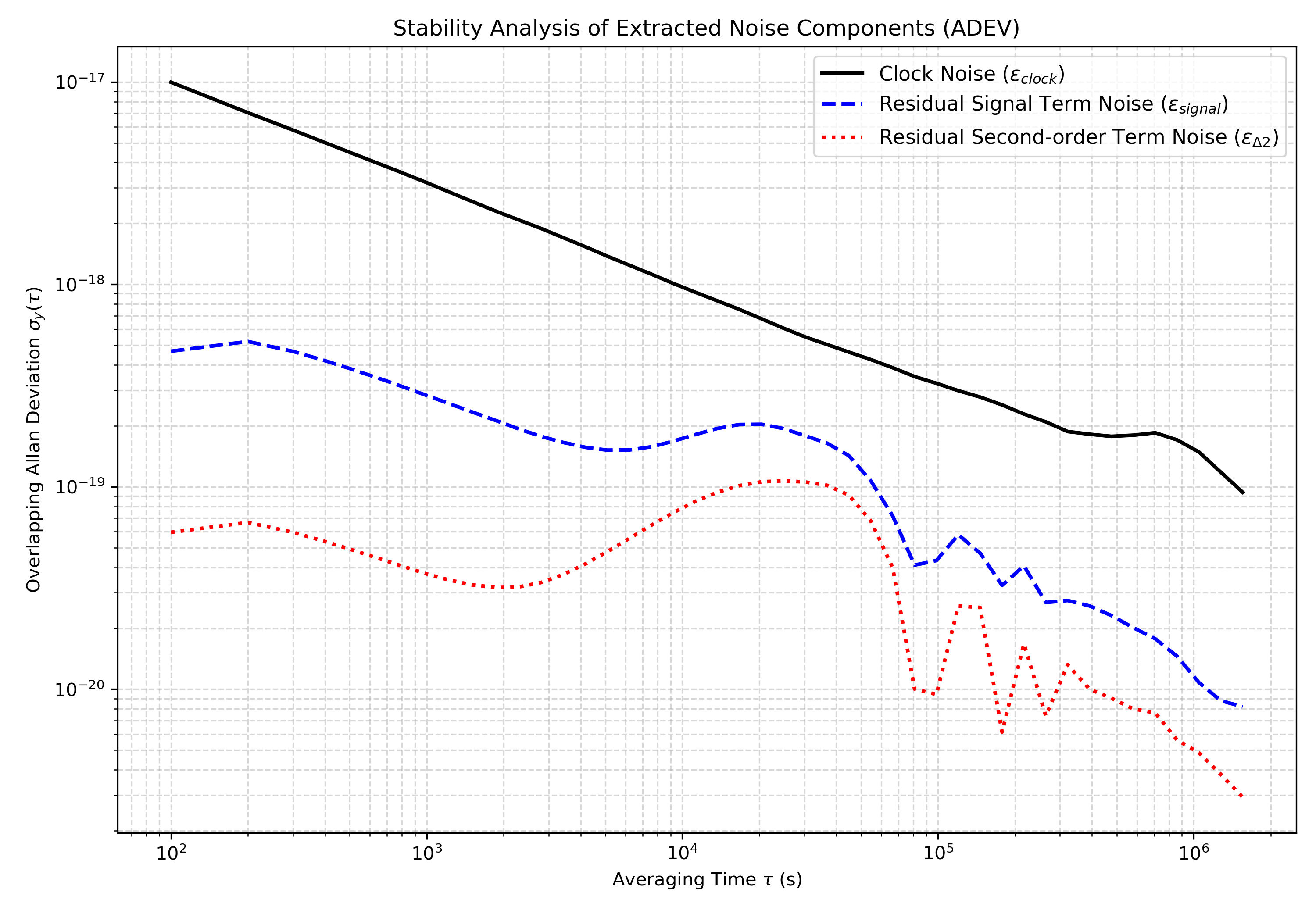}
	\caption{\label{fig:noiseADEV} ADEV of the extracted noise components.}
\end{figure}

After isolating the tidal signal from the dominant nominal relativistic shifts, we characterize the extracted noise components. Figure \ref{fig:noiseseries} presents the time-domain series of the primary error sources: clock noise ($\epsilon_{clock}$), residual signal term noise ($\epsilon_{signal}$), and residual second-order term noise ($\epsilon_{\Delta 2}$). The simulated clock noise dominates the overall amplitude, fluctuating at the $10^{-17}$ level. The frequency-domain stability of these noises is further elucidated through the ADEV plot in Figure \ref{fig:noiseADEV}. The clock noise naturally establishes the fundamental stability floor across all averaging times. Crucially, the residual signal noise ($\epsilon_{signal}$), which is heavily coupled with the satellite's radial orbit errors, exhibits prominent bumps around $\tau \approx 10^4 - 10^5$ seconds. These correspond to the unmodeled 1-CPR and 2-CPR orbital periods. This colored noise signature emphasizes the necessity of utilizing the WLS estimator with a rigorous covariance matrix to prevent parameter biasing.

\begin{figure}
	\includegraphics[width=0.5\textwidth]{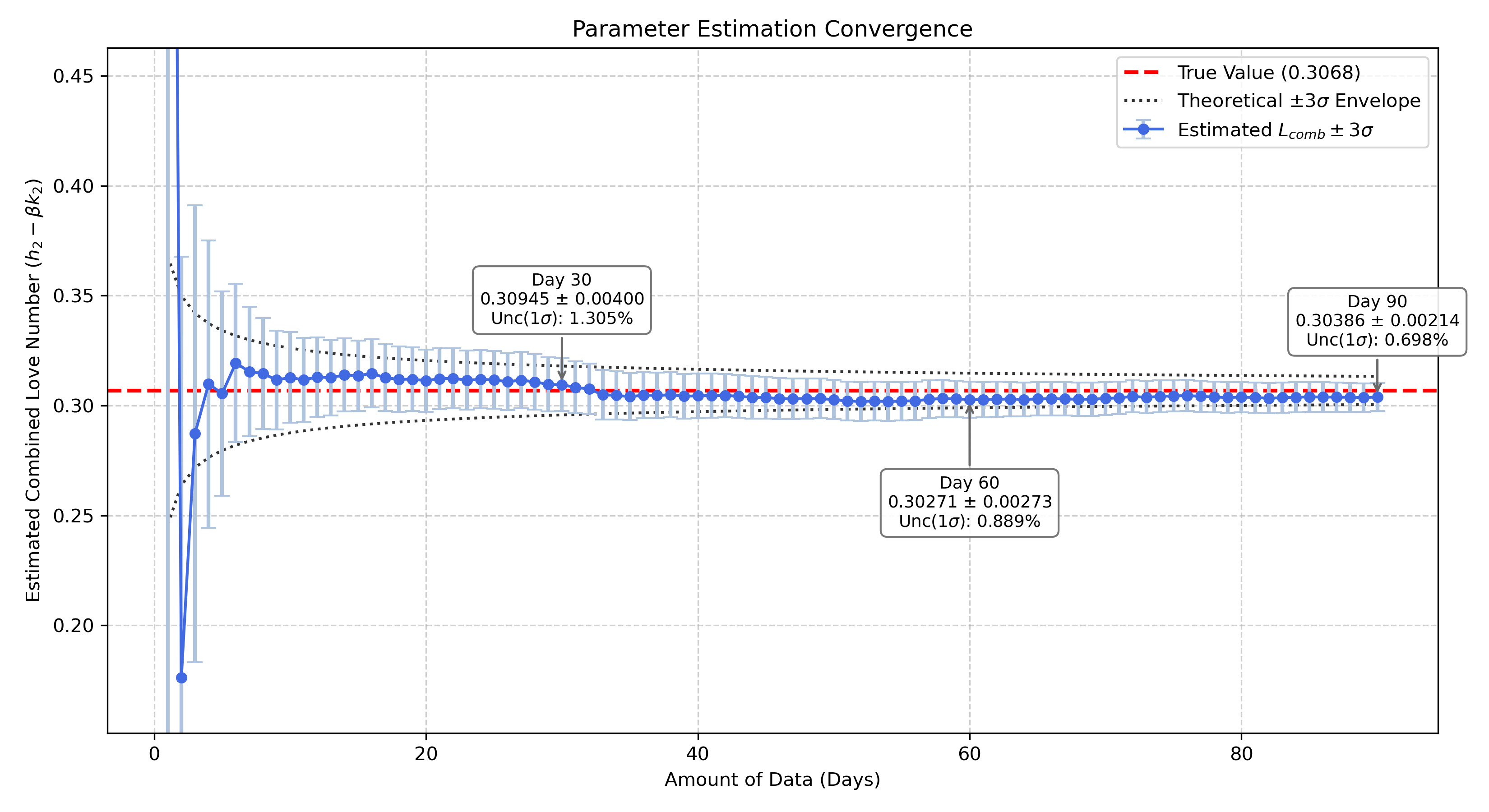}
	\caption{\label{fig:datalengthfit} Convergence of the fitted combined Love number parameter versus observation data length.}
\end{figure}

In single-satellite orbit determination and gravity field recovery, the independent decoupling of the solid Earth Love numbers $h_2$ and $k_2$ is severely hampered by the attenuation of tidal potentials at high IGSO altitudes. Consequently, we parameterize the estimation using the effective combined Love number $L_{\text{comb}} = h_2 - \beta k_2$, with $\beta = 1 - (R_E/r_{s0})^3$. Figure \ref{fig:datalengthfit} demonstrates the convergence history of the estimated $L_{\text{comb}}$ against the continuous data accumulation span. During the initial phase, the estimated values suffer from significant fluctuations and large formal errors, revealing the dominant influence of un-averaged periodic orbital errors and noise. However, with the extension of the observation span, the WLS estimator rapidly converges toward the true value of 0.3068, tightly bounded by the shrinking empirical $3\sigma$ error bars and the theoretical $\pm 3\sigma$ envelope. Specifically, by Day 30, the periodic perturbations are sufficiently suppressed, yielding an estimate of $0.30945 \pm 0.00400$ with a $1\sigma$ relative uncertainty of 1.305\%. Beyond this 30-day threshold, the estimates exhibit asymptotic stabilization, further refining to $0.30386 \pm 0.00214$ (relative uncertainty of 0.698\%) at Day 90. This stable behavior underscores the high fidelity and robustness of the parameter extraction method over extended integration intervals.

\begin{figure}
	\includegraphics[width=0.5\textwidth]{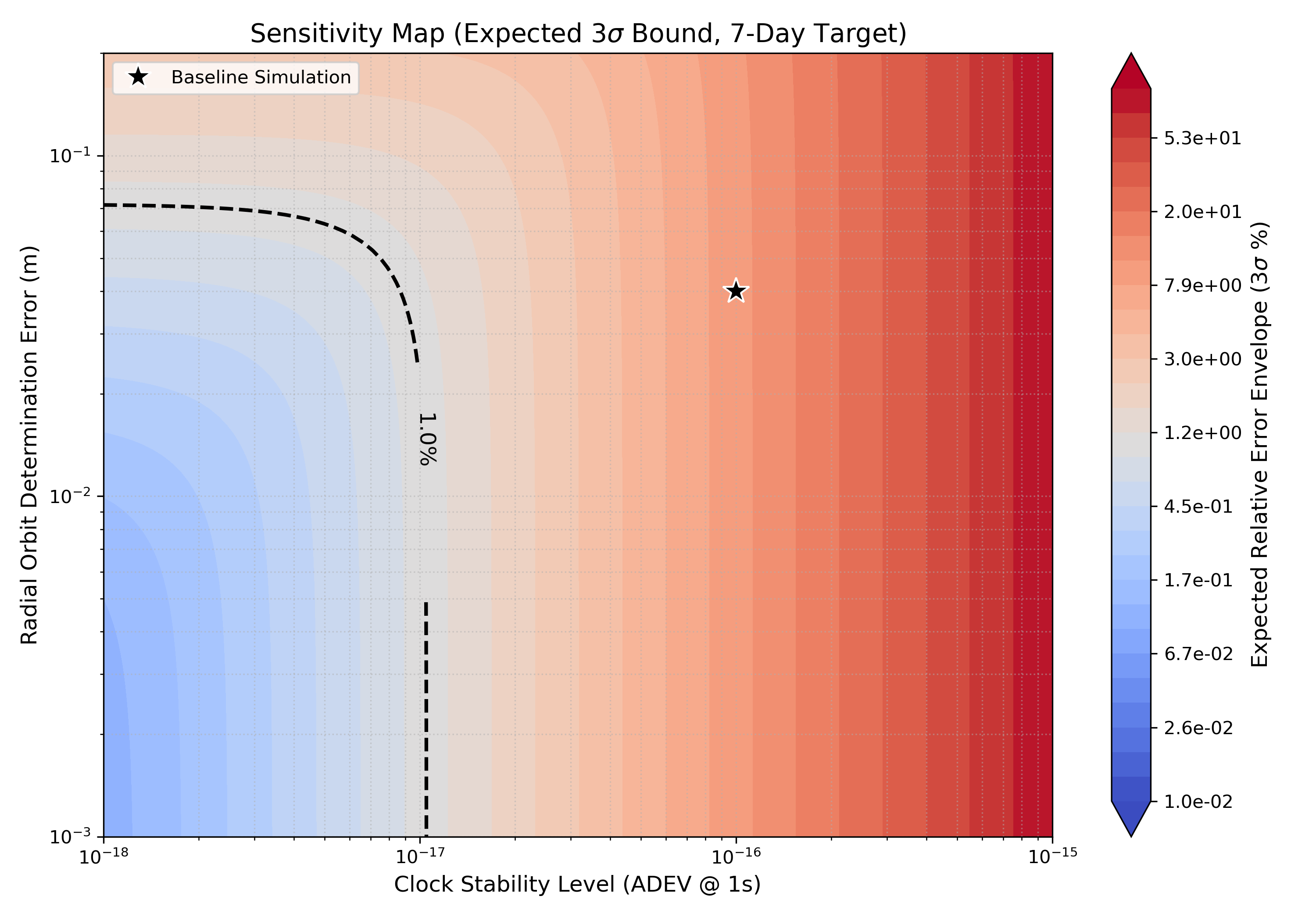}
	\caption{\label{fig:noisesensitivity} Sensitivity map of relative extraction error under varying clock stabilities and orbit determination errors.}
\end{figure}

Finally, to provide a reference for future Relativistic Geodesy missions, we conducted a sensitivity analysis to assess how different physical limitations affect the estimation accuracy. Figure \ref{fig:noisesensitivity} displays a sensitivity map illustrating the relative extraction error of $L_{comb}$ across a parameter space defined by clock stability (ADEV @ 1s) and radial POD error. The baseline simulation of this study is marked by the star symbol. The nearly vertical contours indicate that current estimation accuracy is primarily limited by clock stability rather than POD errors. Consequently, advancing clock stability to the $10^{-19}$ regime will improve geodynamic sensing far more effectively than achieving millimeter-level orbit determination.

\section{CONCLUSION}
\label{conclusion}
In this study, a methodology for estimating solid Earth tide parameters was developed based on an existing relativistic framework for space-to-ground clock comparisons. By incorporating Earth's elastic deformation and direct lunisolar tidal potentials into this framework, analytical expressions were derived that isolate weak degree-2 solid tide signals within a three-link Doppler cancellation scheme. To extract these parameters while accounting for the colored noise of clock instability and unmodeled periodic errors in POD, a WLS estimator was implemented.

A 90-day numerical simulation of an IGSO satellite demonstrated that high-performance optical clock links can successfully extract minute periodic variations in the geopotential. At high orbital altitudes, however, the severe attenuation of the direct tidal potential creates strong collinearity, hindering the independent extraction of the displacement and potential Love numbers ($h_2$ and $k_2$). This limitation was resolved by defining a combined effective parameter ($L_{comb}$), which converges to a stable estimate within a 30-day continuous observation window.

Sensitivity analysis indicates that estimation accuracy is currently limited by clock stability rather than radial POD errors. Consequently, advancing optical clock performance toward higher precision regimes will benefit geodynamic sensing significantly more than achieving millimeter-level orbit accuracy. These results provide a theoretical and methodological basis for utilizing future space-based frequency networks to monitor Earth's internal dynamics.

\section*{Acknowledgements}
This work is supported by the National Key Research and Development Program of China (Grant No. 2023YFC2206100), and the Quantum Science and Technology-National Science and Technology Major Project (Grant No.2021ZD0300106).

\bibliography{solid_tide}

\end{document}